\documentclass[conference]{IEEEtran}
\IEEEoverridecommandlockouts
\usepackage{amsmath,graphicx}
\usepackage[utf8]{inputenc}
\usepackage[english]{babel}
\usepackage{color}
\usepackage{url}
\usepackage{caption}
\usepackage{multirow}
\usepackage{textcomp}
\usepackage{booktabs}
\usepackage{multicol}
\usepackage{algorithm2e}
\usepackage{todonotes}
 \usepackage[nolist,nohyperlinks]{acronym}

\def\x{{\mathbf x}}
\def\L{{\cal L}}
\newcommand{\loss}{\ensuremath{\mathcal{L}}}
\newcommand{\lbce}{\ensuremath{\loss_\text{BCE}}\xspace}
\newcommand{\liou}{\ensuremath{\loss_\text{IoU}}\xspace}
\newcommand{\lDUAL}{\ensuremath{\loss_\text{DUAL}}\xspace}
\newcommand{\glmsrf}{\ensuremath{\loss_\text{GMSRF}}\xspace}
\usepackage[acronym]{glossaries}
\acrodef{mIoU}{mean intersection over union}
\acrodef{DSC}{dice coefficient}

\begin{document}
\title{GMSRF-Net: An improved generalizability with global multi-scale residual fusion network for polyp segmentation}
% Abhishek , Sukulpa , Debesh, Updamada Pal, Sharib Ali
% GIANA - collection of datasets  ( - ETRIS dataset) 
%  Sharib Ali$^{\star}$ %
% \address{${\star}$Institute of Biomedical Engineering, Department of Engineering Science, \\
% University of Oxford, Oxford, UK
\author{\IEEEauthorblockN{Abhishek Srivastava\IEEEauthorrefmark{1},
Sukalpa Chanda\IEEEauthorrefmark{2}, Debesh Jha\IEEEauthorrefmark{3}\IEEEauthorrefmark{4},Umapada Pal\IEEEauthorrefmark{1}, Sharib Ali\IEEEauthorrefmark{5}
}
%\vspace{2mm}
\IEEEauthorblockA{\IEEEauthorrefmark{1}Indian Statistical Institute, India\ \ \ \ \
\IEEEauthorrefmark{2}Østfold University College, Norway\ \ \ \ \ \
\IEEEauthorrefmark{3}SimulaMet, Norway \\
\IEEEauthorrefmark{4}UiT The Arctic University of Norway, Norway \ \ \ \ \ \
\IEEEauthorrefmark{5} University of Oxford,  UK\\
Email: {abhisheksrivastava2397@gmail.com}}}

\maketitle
\def\x{{\mathbf x}}
\def\L{{\cal L}}
\begin{abstract}
Colonoscopy is a gold standard procedure but is highly operator-dependent. Efforts have been made to automate the detection and segmentation of polyps, a precancerous precursor, to effectively minimize missed rate. Widely used computer-aided polyp segmentation systems actuated by encoder-decoder have achieved high performance in terms of accuracy. However, polyp segmentation datasets collected from varied centers can follow different imaging protocols leading to difference in data distribution. As a result, most methods suffer from performance drop and require re-training for each specific dataset. We address this generalizability issue by proposing a global multi-scale residual fusion network (GMSRF-Net). Our proposed network maintains high-resolution representations while performing multi-scale fusion operations for all resolution scales. To further leverage scale information, we design cross multi-scale attention (CMSA) and multi-scale feature selection (MSFS) modules within the GMSRF-Net. The repeated fusion operations gated by CMSA and MSFS demonstrate improved generalizability of the network. Experiments conducted on two different polyp segmentation datasets show that our proposed GMSRF-Net outperforms the previous top-performing state-of-the-art method by 8.34\% and 10.31\% on unseen CVC-ClinicDB and unseen Kvasir-SEG, in terms of dice coefficient.
%Our proposed network maintains high-resolution while simultaneously performing multi-scale fusion operations for all resolution scales
% TODO: Check z and s used uniformly
% generate robust, diverse and representative features that
%However, when gastrointestinal images are saign a mpled from a distribution which differs from the distribution these models are trained  
%In this work we will look into generalisation ability of methods - comparison between multiscale fusion networks to that of other same scale fusion methods - we justify and show experiments how the specific network is more generalisable over other methods -- feature map going to various different paths provide more generalisable segmentation map - etc.
% TODO: Discuss with Sir
% made some changes in abstract
% Added MSFS part in the introduction
% Discuss CMSA part in the method, see the highlighted part.
% In the discussion section, see the highlighted part 0.79 and 0.75 should we specify for which dataset we are getting the performance.
% MOST IMPORTANT: No information about why we are getting better generalization performance using GMSRF, CMSA, MSFS in the discussion section check it.
% Experiments conducted on two different polyp segmentation datasets show that our proposed GMSRF-Net outperforms previous top performing state-of-the-art method by 8.31\% on unseen CVC-ClinicDB and by 10.31\% on unseen Kvasir-SEG in terms of dice coefficient.
% changed this in the abstract
\end{abstract}
\section{Introduction}
\label{sec:intro}
Colorectal cancer (CRC) has been consistently ranked third in terms of prevalence~\cite{howlader2018seer}. The leading cause of CRC is colorectal adenomatous polyps, and thus identification and resection of polyps can reduce the occurrence of CRC. Colonoscopy serves as a gold standard technique for surveillance and treatment. Studies have shown that timely colonoscopy can reduce the chances of CRC by 30\%~\cite{haggar2009colorectal}. However, the success of careful identification of malicious polyps and their subsequent resection depends on the ability and experience of clinicians which makes it prone to human error. Such factors eventually lead to a high polyp missed rate~\cite{puyal2020endoscopic}. Artificial intelligence (AI) driven methods can be effective and provide precise detection and segmentation of polyps. 

With the advent of convolution neural networks (CNNs) research for the polyp segmentation task has been widely conducted to reduce operator-dependent problems in colonoscopy. However, the variations in structures and size of polyps and fluctuation of contrast between polyps and their immediate surrounding make it a challenging task. Whilst methods such as U-Net~\cite{ronneberger2015u}, U-Net++~\cite{zhou2019unet++}, PraNet~\cite{fan2020pranet}, UACA-Net~\cite{kim2021uacanet}, MSRF-Net~\cite{srivastava2021msrf} have demonstrated higher metric performances, when the intervention of imaging protocols varies, performance of these methods fall considerably. The imaging protocols used to acquire colonoscopy images at most times vary over different medical institutions, the performance drop in these methods when tested on unseen data need to be minimized. 

% Related work here:
We can observe various reincarnations of the U-Net developed for polyp segmentation task in~\cite{zhou2019unet++,jha2019resunet++,zhou2018unet++}. Similarly, PraNet~\cite{fan2020pranet} aggregated deep features in their parallel partial decoder to form initial guidance area maps. ColonSegNet~\cite{jha2020real} used only two encoder and two decoder layers that made their network parameters relatively smaller enabling a faster inference time. UACA-Net~\cite{kim2021uacanet} used a saliency map for each level in the decoder to calculate foreground, background, and uncertain area maps.
However, a major drawback with encoder-decoder architectures like U-Net is that shallow features from the encoder and deep features from the decoder suffer from semantic gap~\cite{ibtehaz2020multiresunet}. Deeplabv3+~\cite{chen2018encoder} introduced atrous spatial pyramid pooling with skip connections to aggregate global multi-scale context. Wang et al.~\cite{wang2020deep} designed a network where spatial precision is not compromised by maintaining high-resolution representations throughout the process. Here, multi-scale fusion is performed by repeated cross-scale fusion of features for all resolution scales. Inspired by deep fusion~\cite{sun2018igcv3,xie2018interleaved}, MSRF-Net~\cite{srivastava2021msrf} increased the number of fusion operations by introducing dual-scale dense fusion blocks, which allowed the preservation of both high- and low-level features for all resolution scales. The authors~\cite{srivastava2021msrf} demonstrated the superior generalizability of MSRF-Net and HRNet on polyp segmentation tasks. Building upon these concepts we aim to increase generalizability of polyp segmentation task under different clinical settings by introducing a global multi-scale residual fusion network ``GMSRF-Net''. 

Our GMSRF-Net uses a densely connected multi-scale fusion mechanism that fuses features from all resolution scales at once. The fusion of multi-scale features occurs at each convolutional layer of the densely connected structure which further increases the frequency of fusion operation while maintaining global multi-scale context throughout the process. Additionally, we design a novel cross multi-scale attention (CMSA) mechanism. These attention maps formed by the aggregation of multi-scale context boost the feature map representations in all resolution scales. Our multi-scale feature selection (MSFS) module, applies channel-wise attention on the features fused from all scales to further amplify the relevant features. Experiments demonstrate the improved generalizability of the proposed approach compared to former state-of-the-art (SOTA) methods. Thus, our GMSRF-Net opens new avenues to enhance the generalization capacity of CNN-based supervised learning approaches.

\begin{figure*}[!t]
    \centering
    \includegraphics[height=0.32\textwidth]{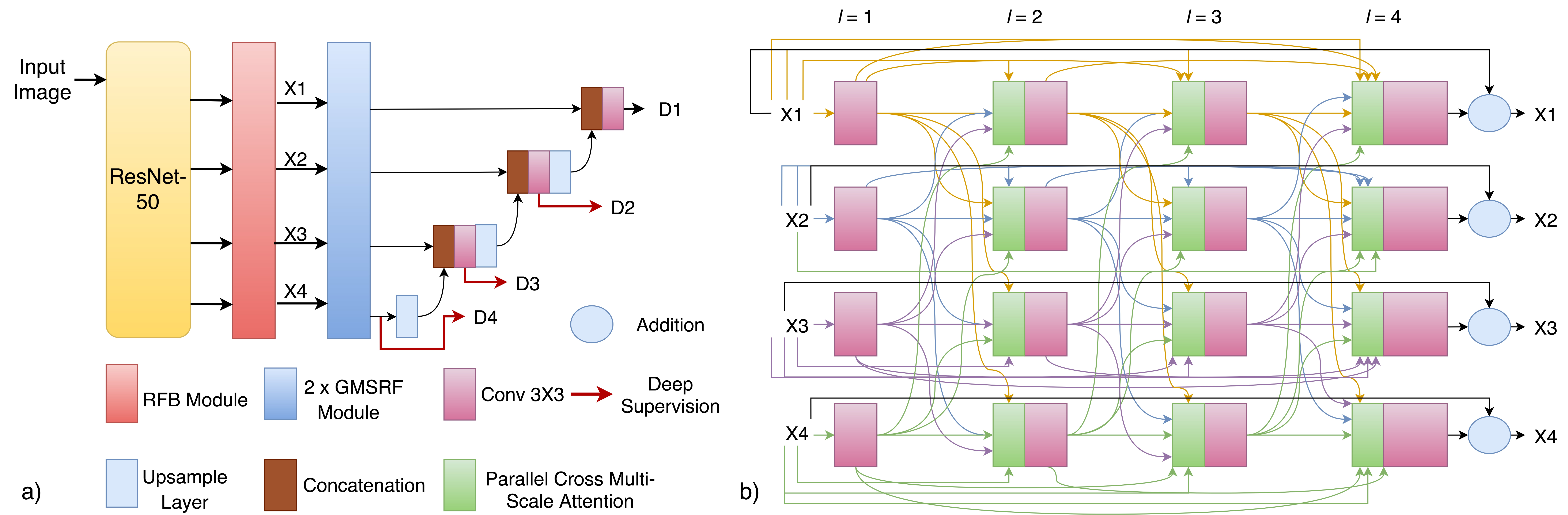}
    \caption{The proposed GMSRF-Net architecture, a) The GMSRF-Net architecture (left), b) The GMSRF module (right)}
    \label{fig:GMSRF-Net}
\end{figure*}
\section{Materials and method}
\subsection{Materials}
We have chosen two standard publicly available polyp segmentation datasets: Kvasir-SEG~\cite{jha2020kvasir} and CVC-ClinicDB~\cite{bernal2015wm}. Kvasir-SEG was acquired in Vestre Viken Health Trust in Norway while CVC-ClinicDB was obtained in Hospital Clinic in Barcelona, Spain. To demonstrate the effectiveness of our technique we perform four experiments with different setups. Two experiments were carried out when the training and testing datasets are the same. Additionally, to establish the generalization capacity of our network, we trained and tested our model on different datasets, i.e., trained on Kvasir-SEG and tested on CVC-ClinicDB and vice versa.
\subsection{Method}
In this section, we present the architecture of our GMSRF-Net (see Fig.~\ref{fig:GMSRF-Net}). GMSRF-Net uses global multi-scale feature fusion mechanism which incorporates cross multi-scale attention and subsequent multi-scale feature selection module for accurate and generalizable segmentation of polyps. The encoder, two GMSRF modules, and decoder are detailed in the following subsections.
\subsubsection{Encoder block}
The colonoscopy images are first processed by ResNet50~\cite{he2016deep} backbone pre-trained on ImageNet. The number of feature maps for all scales is reduced by Receptive Field Blocks (RFBs)~\cite{liu2018receptive} to reduce the computational cost incurred by the following global multi-scale residual fusion (GMSRF) and the decoder network (see Fig.~\ref{fig:GMSRF-Net}(a)). Here, the features generated by the RFB module be denoted as $X_{i}$ where $i \in \{1,2,3,4\}$ denote scales.
% The GMRF module and the parallel cross multi-scale attention and feature selection module entailed within it are described in the following subsection. 
\subsubsection{Global Multi-Scale Residual Fusion block}
Let $[X_{1},X_{2},X_{3},X_{4}]$ denote features of distinct spatial resolutions (see Fig.~\ref{fig:GMSRF-Net}(b)). In the initial layer $l=1$, where $l$ represents the layer number in GMSRF module, each set of feature maps undergoes a convolution operation with output number of feature maps set as $k$, $k$ being the growth factor~\cite{huang2017densely}. 
% We use two GMSRF modules subsequently in our network (see Fig.~\ref{fig:GMSRF-Net}(a)).
% which controls the amount of features generated by each convolution operation in the entire densely connected multi-scale fusion mechanism.

\begin{figure}[!t]
    \centering
    \includegraphics[width=\linewidth]{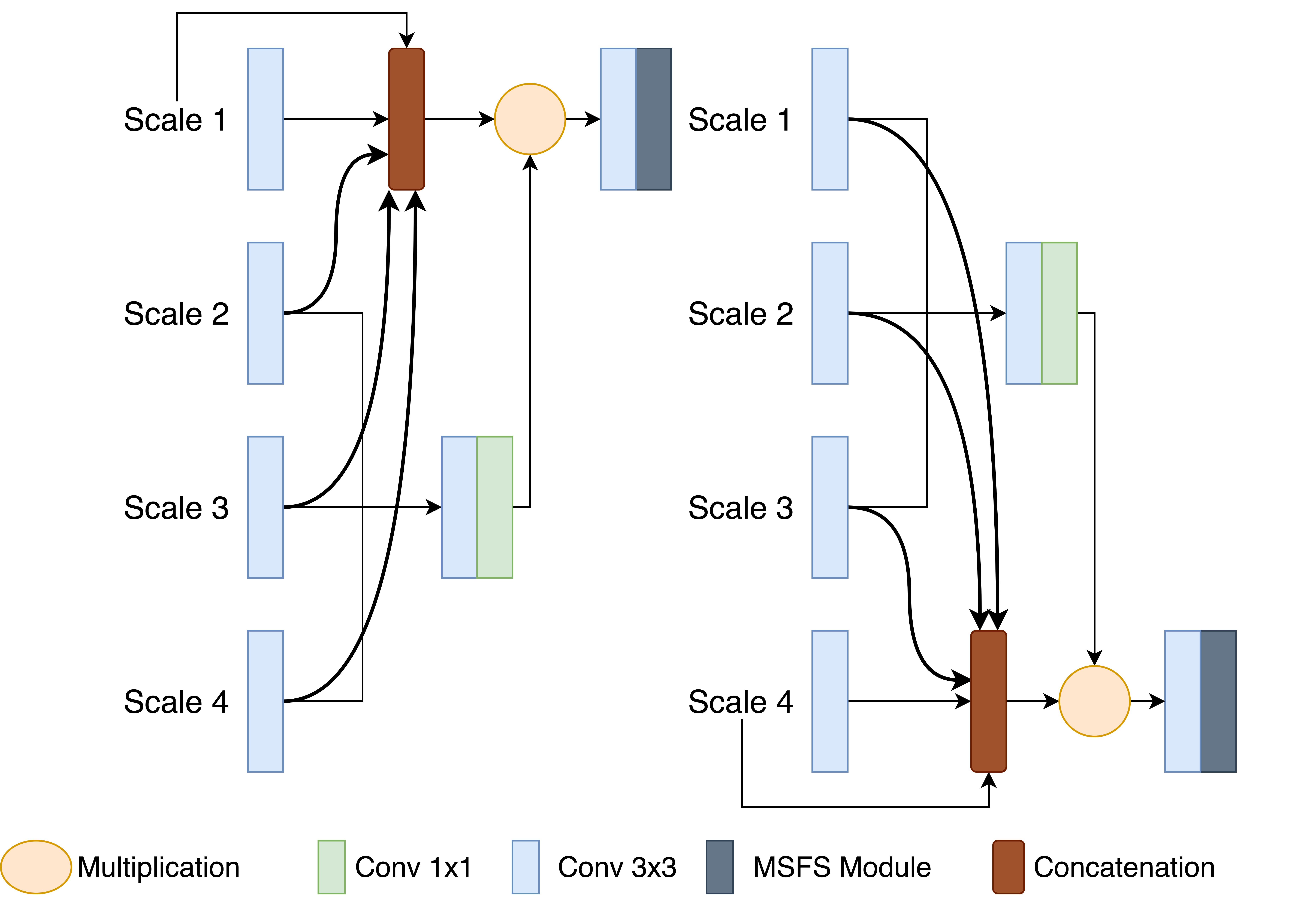}
    \caption{Computation of CMSA and MSFS for scale 1 and scale 4 (computed for scale 2 and scale 3 as well)}
    \label{fig:GMSRF-Attention}
\end{figure}

\textbf{Cross multi-scale attention maps (CMSA)} are calculated for each scale concurrently. Eq.~\ref{eq:crossatt} represents how the $l'th$ CMSA is calculated for the $i'th$ scale. \{$X_{w},X_{y},X_{z}\} \neq X_{i}$ are first transformed to the spatial resolution size of the $i'th$ scale by suitable convolution or transposed convolution operations (see Fig.~\ref{fig:GMSRF-Attention}). They are concatenated and then processed by a $3\times3$ convolution operation, to effectively fuse the features of selected scales. 
\begin{equation}
\begin{split}
        X_{att,\hat{i},l} = Conv_{1\times1}(Conv_{3\times3}(X_{w,l-1} \oplus X_{y,l-1} \\ \oplus X_{z,-1})), \{w, y, z\} \neq i
\end{split}
    \label{eq:crossatt}
\end{equation}
\noindent{Here}, $\oplus$ represents concatenation operation. Attention maps are then generated to identify spatial locations based on the fused multi-scale features of parallel resolution streams. The information conveyed from low-resolution streams helps to boost the feature maps in the high-resolution stream and vice versa. The subsequent combination with the CMSA module allows the selection of features that are relevant towards identifying the region-of-interest.

% CMSA maps are used to identify spatial regions deemed relevant and conducive to generate segmentation maps of high accuracy.

\textbf{Global multi-scale residual fusion (GMSRF)} is performed as described in Eq.~(\ref{eq:gmsf}). The $l'th$ convolutional layer in the $i'th$ resolution stream receives concatenated feature maps from $l-1'th$ convolutional layer from all resolution scales and previous convolutional layers for the same resolution stream (see Fig.~\ref{fig:GMSRF-Net}(b)). This global multi-scale fusion with densely connected blocks increases the number of paths through which feature maps can propagate and undergoes varying operations before contributing to the final segmentation map prediction.
\begin{equation}
\begin{split}
    X_{i,l} = Conv_{3\times3}(X_{i,0} \cdots X_{i,l-1} \oplus X_{w,l-1} \\
    \oplus X_{y,l-1} \oplus X_{z,l-1}) , \{w, y, z\} \neq i
    \end{split}
    \label{eq:gmsf}
\end{equation}
The feature maps can capture the global multi-scale context at each layer of the densely connected mechanism. Eq.~(\ref{eq:attmul}) describes how CMSA maps are used to identify and propagate relevant features of the $i'th$ scale stream forward.
\begin{equation}
    X_{i,l} = X_{i,l} \otimes X_{att,\hat{i,l}}
    \label{eq:attmul}
\end{equation}
%
% For Abhishek TODO: there is a paper which states high res features are not as important for performance find it and use it in the following paragraph 
\textbf{Multi-scale feature selection (MSFS)} module, is the next step where channel-wise attention is applied on the fused features using squeeze and excitation ($\text{S\&E}$) block~\cite{hu2018squeeze} (refer to Fig.~\ref{fig:GMSRF-Attention}). This enables the amplification of salient channels transmitted by various scale streams. The suppression of irrelevant channels by this module is also conducive to a higher level of accuracy.
% To further realize the potential of this technique, the last layer of the GMSRF module i.e $l=4$, we do not reduce the number of output channels by $1x1$ convolution before applying squeeze and excitation block. 
Residual connection from the input of the GMSRF module is added to improve gradient flow. For simplification purposes, we use the $i'th$ scale while describing this mechanism. 
% Generation of CMSA, GMSRF, and MSFS for all scales i.e $i,w,y,z$ are being performed concurrently.
%
%
%\vspace{-2.00mm}
\subsubsection{Decoder}
To fully establish the contribution of our GMSRF-Net, we choose to use a vanilla decoder (see Fig.~\ref{fig:GMSRF-Net}(a)). $X_{i}$ is the output of the GMSRF module for the $i'th$ scale. Each decoder block upscales the output from the previous decoder block and concatenates the resultant feature maps from the same scale output of the GMSRF module (see Equation~\ref{eq:decoder}).
\begin{equation}
   D_{i} = Conv_{3\times3}(TransConv(D_{i-1}) \oplus X_{i})
   \label{eq:decoder}
\end{equation}
\noindent{Here}, TransConv is strided transposed convolutional layer and initially $D_{4} = X_{4}$. The output of all decoder blocks, i.e $D_{4},D_{3},D_{2}$, are upscaled to the size of ground truth maps for improved gradient flow and regularization. 
% We have not used any established attention based mechanism to display that GMSRF module only aided by ResNet50 backbone and RFB's is able to display comparable results to state-of-the-art MSRF-Net~\cite{srivastava2021msrf} on seen datasets and superior results on unseen datasets even under a supervised setting.

\subsubsection{Loss Function}
We use a dual loss function $ \lDUAL = \lbce + \liou$, where $\lDUAL$ is a combination of weighted intersection over union loss (\liou) and binary cross entropy (\lbce). For all supervise segmentation maps generated by all decoder levels, the total loss function is given by: $ \glmsrf = \sum_{i=1}^{i=4} \lDUAL(D_{i})$, where $i$ is the number of decoder layers.

%
% \begin{equation}
%     \lDUAL = \lbce + \liou
%     \label{eq:dualloss}
% \end{equation}
%
% We use the dual loss function to supervise segmentation maps generated by all decoder levels, the total loss function is described in Equation~\ref{eq:gmsrfloss}.
%
% \begin{equation}
%     \glmsrf = \sum_{i=1}^{i=4} \lDUAL(D_{i})
%     \label{eq:gmsrfloss}
% \end{equation}
%
\begin{table*}[!t]
\centering
%\scriptsize
\footnotesize
\caption{Result comparison when source dataset is Kvasir-SEG}
\begin{tabular}{@{}l|l|l|l|l|l|l|l|l@{}}
\toprule
\multirow{2}{*}{\textbf{Method}} & \multicolumn{4}{|c|}{\bf Source data} & \multicolumn{4}{|c}{\bf Unseen dataset ``CVC-ClinicDB''} \\ 
& \textbf{DSC} & \textbf{mIoU} & \textbf{Recall} & \textbf{Precision} & \textbf{DSC} & \textbf{mIoU} & \textbf{Recall} & \textbf{Precision} \\
\hline
U-Net~\cite{ronneberger2015u} & 0.8629  & 0.8176 & 0.9094  &0.8901 & 0.7172 & 0.6133 & 0.7255 &0.7986  \\ \hline
U-Net++~\cite{zhou2019unet++} & 0.7475 & 0.6313 & 0.6865 & 0.8871 & 0.4265  & 0.3345 & 0.3939 & 0.6894 \\ \hline
% ResUNet++~\cite{jha2019resunet++} & 0.8189 & 0.7918 & 0.8372 & 0.9255\\ \hline
Deeplabv3+( Xception)~\cite{chen2018encoder} & 0.8965 & 0.8575 & 0.8984 & 0.9496 & 0.6509 & 0.5385 & 0.6251 & 0.7947 \\ \hline

Deeplabv3+ (Mobilenet)~\cite{chen2018encoder} & 0.8656 & 0.8186 & 0.8808 & 0.9205 & 0.6303 & 0.4825 & 0.5957 & 0.7173 \\ \hline
% {DoubleUNet}~\cite{jha2020doubleu} &0.8699 & 0.8166 &0.9039 & 0.8942\\ \hline
HRNetV2-W18-Smallv2~\cite{Wang_2020} & 0.8179 & 0.7470 & 0.8016 & 0.8696 & 0.6428 & 0.5513 & 0.6811 & 0.7253\\ \hline
HRNetV2-W48~\cite{Wang_2020} & 0.8896 & 0.8262 & 0.8973 & 0.9056 & 0.7901 & 0.6953 & 0.8796 & 0.7694\\ \hline
ColonSegNet~\cite{jha2020real} & 0.8203 & 0.7435 & 0.8124 & 0.8832 & 0.6895 & 0.5813 & 0.7862 & 0.7177 \\ \hline
% DDANet~\cite{tomar2020ddanet}& 0.8915 & 0.8393 & 0.8927 & 0.9213 \\ \hline
% ResUNet++ + CRF$^\diamond$~\cite{jha2021comprehensive}&0.7965 & 0.8250 &0.8119 &0.8045 \\ \hline
PraNet~\cite{fan2020pranet}  & 0.9078 & 0.8561 & 0.9034 & 0.9352 & 0.7225 & 0.6328 & 0.7531 & 0.7888 \\ \hline
UACANet-S~\cite{kim2021uacanet} & 0.8800 & 0.8250 & 0.8701 & 0.9229 & 0.5683 & 0.4907 & 0.5792 & 0.7095\\ \hline
UACANet-L~\cite{kim2021uacanet} & 0.9014 & 0.8555 & 0.8897 & 0.9381 & 0.5589 & 0.4849 & 0.5800 & 0.6775\\ \hline
MSRF-Net~\cite{srivastava2021msrf} & 0.9217 & \bf {0.8914} & 0.9198 & \textbf{0.9666} & 0.7921 & 0.6498 & 0.9001 & 0.7000\\ \hline
GMSRF-Net & \bf {0.9263} & 0.8843 & \bf {0.9402} & 0.9310 & \bf {0.8755} & \bf {0.8091} & \bf {0.9106} & \bf {0.8588} \\ \hline
\bottomrule
\end{tabular}
\label{tab:result1}
\end{table*}

\begin{table*}[!t]
\centering
%\scriptsize
\footnotesize
\caption{Result comparison when source dataset is CVC-ClinicDB}
\begin{tabular}{@{}l|l|l|l|l|l|l|l|l@{}}
\toprule
\multirow{2}{*}{\textbf{Method}} & \multicolumn{4}{|c}{\bf Source data} & \multicolumn{4}{|c}{\bf Unseen dataset ``Kvasir-SEG''} \\ 
& \textbf{DSC} & \textbf{mIoU} & \textbf{Recall} & \textbf{Precision} & \textbf{DSC} & \textbf{mIoU} & \textbf{Recall} & \textbf{Precision} \\
\hline
U-Net~\cite{ronneberger2015u}  & 0.9145 & 0.8654 &0.9178 &0.9381 & 0.6222 & 0.4588 & 0.5129 & 0.8133\\ \hline

U-Net++~\cite{zhou2019unet++} & 0.8453 & 0.7559 & 0.8917 & 0.8323 & 0.5926 & 0.4564 & 0.7352 & 0.5462\\ \hline

Deeplabv3+ (Xception)~\cite{chen2018encoder} & 0.8897 & 0.8706 & 0.9251 & 0.9366 & 0.6746 & 0.5327 & 0.7757 & 0.6296\\\hline

Deeplabv3+ (Mobilenet)~\cite{chen2018encoder} & 0.8985 & 0.8588 & 0.9160 & 0.9287 & 0.6474 & 0.5098 & 0.6632 & 0.6878 \\ \hline

HRNetV2-W18-Smallv2~\cite{Wang_2020} & 0.9073 & 0.8457 & 0.9137 & 0.9191 & 0.7012 & 0.6009 & 0.7184 & 0.7666 \\ \hline 

HRNetV2-W48~\cite{Wang_2020} & 0.9244 & 0.8747 & 0.9234 & 0.9296 & 0.7404 & 0.6233 & 0.7293 & 0.8511\\ \hline

ColonSegNet~\cite{jha2020real} & 0.9132 & 0.8600 & 0.9072 & 0.9292 & 0.6324 & 0.5183 & 0.6112 & 0.7897 \\ \hline

PraNet~\cite{fan2020pranet}  & 0.9072 & 0.8575 & 0.9227 & 0.9134 & 0.7293 & 0.6262 & 0.8007 & 0.7623\\ \hline

UACANet-S~\cite{kim2021uacanet} & 0.9190 & 0.8700 & 0.9285 & 0.9201 & 0.6945 & 0.5894 & 0.7692 & 0.7377\\ \hline

UACANet-L~\cite{kim2021uacanet}& 0.9098 & 0.8649 & 0.9174  & 0.9114 & 0.7312  & 0.6383 & 0.7417 & 0.8314\\ \hline

MSRF-Net~\cite{srivastava2021msrf} & \textbf{0.9420} & \textbf{0.9043} & \textbf{0.9567} & \textbf{0.9427} & 0.7575 & 0.6337 &0.7197 &0.8414 \\ \hline
GMSRF-Net & 0.9326 & 0.8882 &  0.9376 & 0.9307 & \textbf{0.8606} & \textbf{0.7877} & \textbf{0.8641} & \textbf{0.9056} \\ \hline
\bottomrule
\end{tabular}
\label{tab:result2}
%\vspace{-5mm}
\end{table*}

\begin{figure}[!t]
    \centering
    \includegraphics[width=\linewidth]{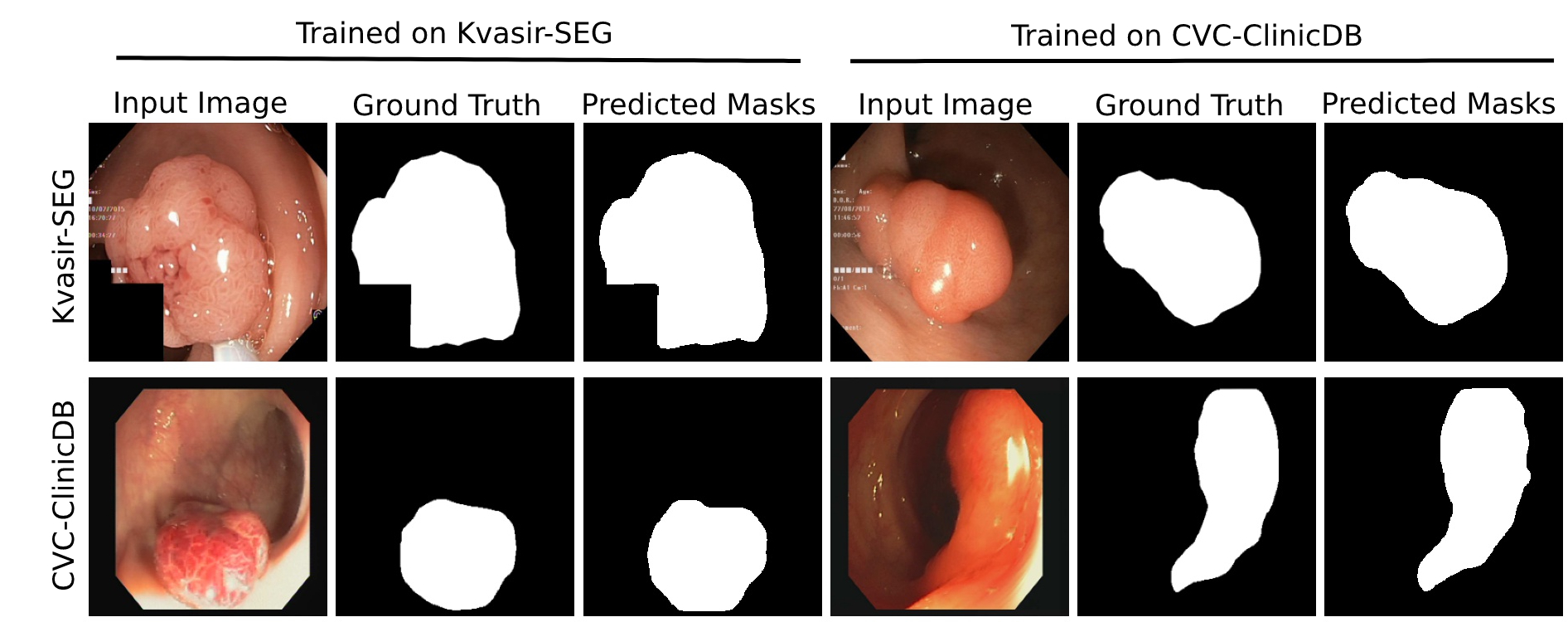}
    \caption{Qualitative results for GMSRF-Net}
    \label{fig:qualitative}
\end{figure}
\section{Experiments}
\subsection{Experimental setup}
We evaluate our proposed GMSRF-Net on Kvasir-SEG and CVC-ClinicDB. All images are resized to $256\times256$ as a pre-processing step. We reserve 80\% data for training, 10\% for validation, and 10\% for testing. The training set is augmented using techniques like random flipping, cropping, color jittering etc. All experiments are conducted using the same train-val-test configuration. We train the network for 50 epochs using Adam optimizer with initial learning rate of $1e-4$ and batch size of 8. All experiments were performed on an NVIDIA DGX-2 machine using NVIDIA V100. 

\subsection{Results and discussion}
From Table~\ref{tab:result1}, it can be observed that our GMSRF-Net is competitive to MSRF-Net on the same source data (Kvasir-SEG), while outperforming on unseen data (CVC-ClinicDB). An increase of 8.34\%, 15.93\%, 1.05\%, 15.88\% in \ac{DSC}, \ac{mIoU}, recall and precision, respectively, can be seen when compared with the best performing SOTA method (MSRF-Net). A similar trend can be noted in Table~\ref{tab:result2} where our proposed method outperformed SOTA method on unseen Kvasir-SEG by large margins on all metrics: improvement of 10.31\%, 15.40\%, 14.44\% and 6.42\% on \ac{DSC}, \ac{mIoU}, recall and precision, respectively. Moreover, we can see that GMSRF-Net achieves a \ac{DSC} of 0.8606 when trained on CVC-ClinicDB and tested on Kvasir-SEG (see Table~\ref{tab:result2}). However, some networks such as U-Net++, ColonSegNet, and HRNetV2-W18-Smallv2 reports relatively lower performance even when they are trained and tested on Kvasir-SEG (see Table~\ref{tab:result2}). Fig.~\ref{fig:qualitative} exhibits that our GMSRF-Net produces optimal segmentation mask predictions in both scenarios when training and test sources are either the same or different.

Our experiments demonstrate that the multi-scale fusion technique that combines features from all resolution scales such as HR-Net and MSRF-Net yields better generalization performances (see Table~\ref{tab:result1}-\ref{tab:result2}). Our GMSRF-Net using global multi-scale residual fusion increases the number of fusion operations together with attention modules (CMSA and MSFS) achieving an improved generalization ability. 

\section{CONCLUSION}
% In this paper, we propose a global multi-scale fusion technique which incorporates CMSA and MSFS mechanism for aggregating global context at each stage, maintenance of high resolution representations and feature enrichment of high resolution feature stream using features from low resolution features and vice versa. 
In this paper, we propose a global multi-scale feature fusion technique that incorporates CMSA and MSFS mechanisms for aggregating reliable and robust global features at each stage. Our proposed network maintains high resolution representations and enriches high-resolution features by fusion with low-resolution feature streams and vice versa.~The proposed technique achieves significant performance gain on segmentation tasks where the training and testing datasets are from different distributions.~The generalization performance of our GMSRF-Net is an important step towards improving the generalizability of supervised learning methods.~In future, we will extend our work towards quantifying the generalizability of the proposed model on other biomedical imaging datasets. 
%The generalization performance of GMSRF-Net establishes that deep-learning based segmentation can perform effectively on unseen data distributions even under a supervised setting. In future, we will extend our work to other biomedical classification and segmentation tasks.

\small{\subsubsection*{COMPLIANCE WITH ETHICAL STANDARDS}} This research study was conducted retrospectively using human subject data made available in open access by Kvasir-SEG and CVC-ClinicDB. Ethical approval was not required.
%% \small{\subsubsection*{ACKNOWLEDGEMENTS}}
%\vfill
%\pagebreak
\bibliographystyle{IEEEtran}
\bibliography{refs}
\end{document}